\documentclass{article}
\usepackage{spconf,amsmath,graphicx,svg}

\usepackage{adjustbox}
\title{TWO-STAGE DOMAIN ADAPTED TRAINING FOR BETTER GENERALIZATION \\  IN REAL-WORLD IMAGE RESTORATION AND SUPER-RESOLUTION}
\name{Cansu Korkmaz, A.Murat Tekalp\sthanks{This work was supported by TUBITAK projects 217E033 and 120C156, and a grant from Turkish Is Bank to KUIS AI Center. A. M. Tekalp also acknowledges support from Turkish Academy of Sciences (TUBA).}, Zafer Do\u{g}an\sthanks{Z.D. acknowledges that this work was partially supported by the TUBITAK 2232 International Fellowship for Outstanding Researchers Award (No. 118C337) and an AI Fellowship provided by the KUIS AI Lab.}}
\address{Koc University, Dept. of Electrical and Electronics Engineering, Istanbul, Turkey \\
\{ckorkmaz14, mtekalp, zdogan\}@ku.edu.tr}
\begin{document}
%
\maketitle
\begin{abstract}
It is well-known that in inverse problems, end-to-end trained networks overfit the degradation model seen in the training set, i.e., they do not generalize to other types of degradations well. Recently, an approach to first map images downsampled by unknown filters to bicubicly downsampled look-alike images was proposed to successfully super-resolve such images. In this paper, we show that any inverse problem can be formulated by first mapping the input degraded images to an intermediate domain, and then training a second network to form output images from these intermediate images. Furthermore, the best intermediate domain may vary according to the task. Our experimental results demonstrate that this two-stage domain-adapted training strategy does not only achieve better results on a given class of unknown degradations but can also generalize to other unseen classes of degradations better.   
\end{abstract}

\vspace{-2mm}
\section{Introduction}
\label{sec:intro}

Real-world image restoration and super-resolution (SR) aim to recover enhanced and high-resolution images from their degraded and low-resolution (LR) counterparts, respectively. Most existing learning-based methods rely on the availability of such image pairs modeling realistic image degradations. However, there are two main challenges in practical scenarios: (i) there are only a limited number of  degraded - ground-truth (GT) image pairs with real-world actual degradations; (ii) even if such  dataset is available, the degradation model between different pairs of degraded and GT images may differ significantly.
\begin{figure}[htb]
\begin{minipage}[b]{0.72\linewidth}
  \centering
  \centerline{\includegraphics[width=6cm]{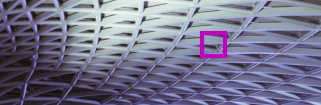}} \vspace{-3pt}
  \centerline{(a)}
\end{minipage}
\begin{minipage}[b]{.24\linewidth}
  \centering
  \centerline{\includegraphics[width=2.0cm]{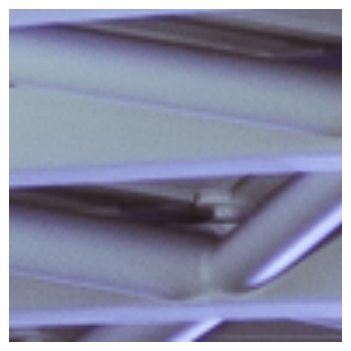}} \vspace{-3pt}
  \centerline{(b)}
\end{minipage} 
\begin{minipage}[b]{.24\linewidth}
  \centering
  \centerline{\includegraphics[width=2.0cm]{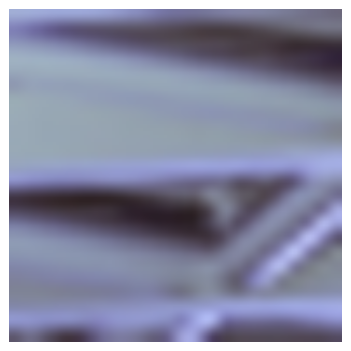}} \vspace{-3pt}
  \centerline{(c)}\medskip
\end{minipage}
\hfill
\begin{minipage}[b]{0.24\linewidth}
  \centering
  \centerline{\includegraphics[width=2.0cm]{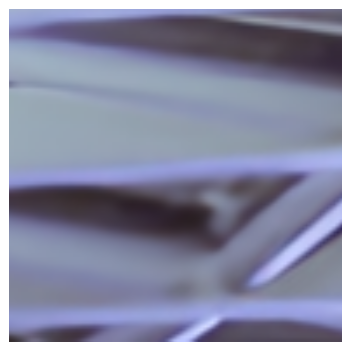}}\vspace{-3pt}
  \centerline{(d)}\medskip
\end{minipage}
\begin{minipage}[b]{.24\linewidth}
  \centering
  \centerline{\includegraphics[width=2.0cm]{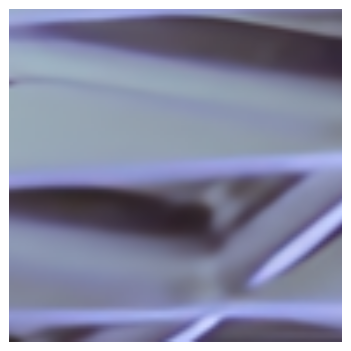}}\vspace{-3pt}
  \centerline{(e)}\medskip
\end{minipage}
\hfill
\begin{minipage}[b]{0.24\linewidth}
  \centering
  \centerline{\includegraphics[width=2.0cm]{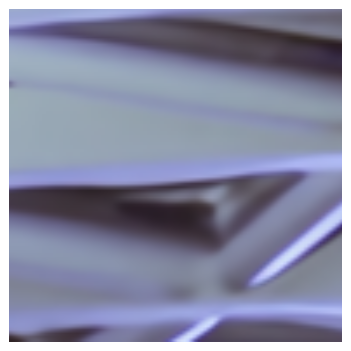}}\vspace{-3pt}
  \centerline{(f)}\medskip
\end{minipage}  \vspace{-15pt}
\caption{This example demonstrates that $\times$4 SR by two-stage domain adapted training not only yields higher PSNR, but also better visual results. (a) HR image DIV2K 0892, (b)~cropped HR, (c) bicubic~interpolation~of~$\downarrow$4~image -25.50dB, (d) two-stage training: \textit{Mapping} and $\times$4 SR -28.97dB, (e) direct $\times$4 SR -29.51dB,  (f) two-stage training: \textit{Mapping$\times$2} and $\times$2 SR -30.02dB. Note that \textit{Mapping} and \textit{Mapping$\times$2} map to two different intermediate domains (see~Fig.~\ref{fig:mapping_diagram}) and \textit{Mapping$\times$2} in (f) gives the best results. 
}
\label{fig:results}\vspace{-2mm}
\end{figure}

\begin{figure*}[htb]
  \centering
  \includegraphics[width=0.9\textwidth]{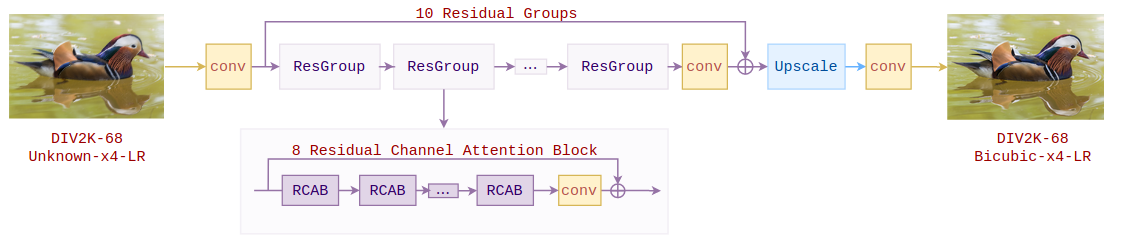}
\vspace{-17pt}
\caption{RCAN architecture \cite{RCAN2018} is used for domain adaptation.}
\label{fig:mapping_architecture}
\end{figure*}

As a result, most learned image restoration and SR methods rely on synthetically generated degraded and GT image pairs to train a network in a fully supervised manner, where degraded images are generated from GT images using a specific degradation model, e.g., low resolution (LR) images are obtained from GT images using the bicubic downsampling. However, this common strategy of training restoration/SR networks on synthetic datasets does not generalize well to real-world image restoration and SR, resulting in visible artifacts.
We refer to this phenomenon as overfitting the degradation model as the learned restoration/SR network overfits the degradation model used to generate the synthetic training set, while the kinds of degradations encountered in real images differ from the assumed degradation model. 

There have been many recent works and challenges with datasets designed for real-world image SR. For example, the RealSR dataset \cite{cai2019toward} was designed for NTIRE 2019 \cite{caiNTIRE2019} and AIM 2019 challenges. 
Our work has been inspired by a recent paper on the conversion of real LR images to bicubic look-alike ones using a GAN-based architecture \cite{Rad2021}. We provide a review of related work and our contributions in Sec.~\ref{sec:related}. Sec.~\ref{sec:method} introduces our methodology. Experimental results are presented in Sec.~\ref{sec:experiment}. Finally, Section~\ref{sec:conclude} concludes the paper.

\section{Related Work and Contributions}
\label{sec:related} \vspace{-6pt}
Image restoration/SR networks have significantly outperformed conventional methods when trained and tested with a known degradation model~\cite{dong2015image, Schocher2018, Kernel}. 
Examples of leading SR networks include Enhanced Deep Super-resolution (EDSR)~\cite{EDSR2017} that optimizes SRResNet \cite{ledig2017photorealistic} and Deep Residual Channel Attention Networks~(RCAN)~\cite{RCAN2018} that introduces residual in residual structure and a channel attention block. 

When it comes to real-world problems, there are a few alternatives. Recently, the RBSR method~\cite{Rad2021} proposed handling real-world SR problem in two steps: first, converting real-world LR images to bicubic look-alike ones, and second, inputting them to an SR network trained on bicubic LR images. Although our approach has common features with RBSR~\cite{Rad2021}, there are some important differences: First,
RBSR generates bicubic look-alike at the same resolution as the input real-world image. In contrast, we propose alternative intermediate domains, possibly at different resolution. For example,
for $\times$4 SR, we propose to generate an intermediate bicubic look-alike image at $\times$2 resolution. This way, the second training stage can serve to correct artifacts that may be generated in the first $\times$2  SR stage.
Moreover, RBSR uses a generative adversarial network to create bicubic look-alike images, which also generates some unwanted textures on bicubic look-alike images. They propose a bicubic perceptual loss to avoid this problem. In contrast, we use RCAN \cite{RCAN2018} for domain mapping with $l_1$-loss that do not encounter unwanted texture problem, and can be trained faster than GAN-based alternatives.  



The main contributions of this paper are threefold: first, we use RCAN to map input degraded images to an intermediate domain; thus, avoid unwanted texture problem
and, we show that two-stage domain adapted training improves performance even in the well-studied $\times$4 bicubic SR; second, we show that two-stage domain mapping $\times$2 performs better than two-stage domain mapping -keeps spatial dimensions same for unknown and mapped bicubic images- and also performs better than the direct training, and we provide quantitative and qualitative results to further show that two-stage domain mapping $\times$2 generalizes better than the other two methods; and finally, we show that two-stage adapted training by mapping input images to an intermediate domain can be applied to other real-world inverse problems, such as image restoration.

\section{Methodology}
\label{sec:method} \vspace{-3pt}
This section presents the proposed two-stage domain-adapted training for blind SR and image restoration problems.
\vspace{-6pt}
 
\subsection{Domain Mapping}
\label{sec:map}
The first stage of the proposed two-stage domain adapted training strategy is a mapping from a target domain (unknown degradation) to a range domain (known degradation). Specifically for the SR task, from unknown low-resolution (LR) images to the domain of bicubic downsampled images.

Domain mapping is performed by RCAN architecture \cite{RCAN2018}, with 10 residual groups and 8 residual channel attention blocks within each residual group, as depicted in Fig. \ref{fig:mapping_architecture}.

\subsection{Application I: $\times$4 Super-Resolution}
\label{sr}
\begin{figure}[htb!]
  \centering
  \includegraphics[width=0.45\textwidth]{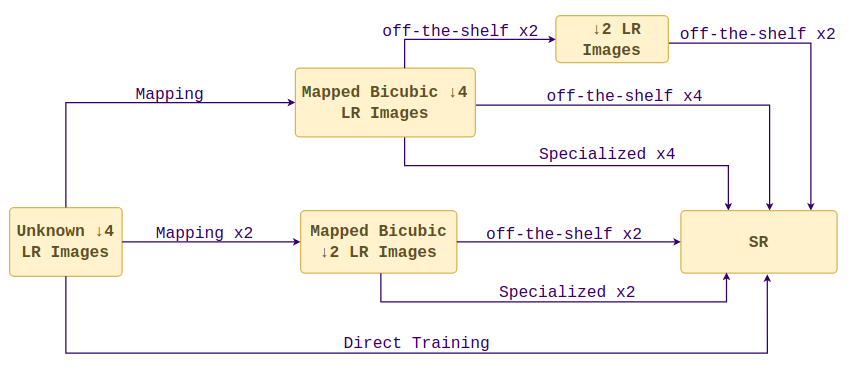}
\vspace{-12pt}
\caption{Possible intermediate domains for $\times$4 SR. ``Mapping" and ``Mapping $\times$2" refer to domain mapping at the same resolution and with $\times$2 increase in resolution, respectively. ``Off-the-shelf $\times$2" refers to a network pre-trained on DIV2K \cite{Timofte_2018_CVPR_Workshops} $\downarrow$2 bicubicly downsampled  $\to$ GT and ``Off-the-shelf $\times$4" refers to a network pre-trained on $\downarrow$4 bicubicly downsampled $\to$ GT, where $\to$ denotes input to output image pairs.}
\label{fig:mapping_diagram}
\end{figure}

In the case of $\times$4 SR, we have two candidates for the intermediate domain, mapping to bicubic look-alike images at the same resolution as the input image (top branch), or mapping to bicubic look-alike images at $\times$2 resolution (middle branch) as depicted in Fig.~\ref{fig:mapping_diagram}. The bottom branch is the classic method that does not involve mapping to an intermediate domain.
\vspace{-3mm}

Following the top branch, there are three options to process the intermediate domain images:
i) apply a pre-trained  $\times$2 SR network back-to-back twice,
ii) apply a pre-trained  $\times$4 SR network,
iii) train a specialized  $\times$4 SR network given the mapped intermediate domain images and GT images. 

Following the middle branch, we can process intermediate domain images by either applying an off-the-shelf $\times$2 SR network or train a specialized  $\times$2 SR network given the mapped intermediate domain images and GT images. Experimental results in Sec.~\ref{sec:experiment} demonstrate that two-stage training with an adapted intermediate domain $\times$2 followed by specialized  $\times$2 SR network always yields the best results and best generalization performance to other test datasets consisting of real-world images. 
\vspace{-6pt}

\subsection{Application II: Image Restoration - Unknown Blur}
\label{restoration}

We apply our method of two-stage domain-adapted training to the real-world image restoration problem with unknown blur. In this case, input blurred images are first mapped to an intermediate domain of a pre-determined blur using the same RCAN architecture \cite{RCAN2018}, without the upsampling layer. These intermediate images are restored by a network trained on blurred (with the pre-determined blur) - GT image pairs.

To the best of our knowledge, this is the first paper that shows the advantages of two-step domian-adapted training for SR and image restoration problems to overcome overfitting the degradation model in real-world inverse problems.

\vspace{-2mm}
\begin{table}[b]
    \centering
    \begin{adjustbox}{width=0.4\textwidth}
    \begin{tabular}{|c|c|c|c|}
     \hline
      & x4 SR & x2x2 SR & Specialized x2x2 SR \\
    \hline
    PSNR(dB) & 28.907 & 28.942 & \textbf{28.967}\\
    \hline
    SSIM & 0.837 & 0.824 & \textbf{0.843} \\
    \hline
    LPIPS-Alex & 0.2741 & 0.2740 & \textbf{0.2701}\\
    \hline
    LPIPS-VGG & 0.3059 & \textbf{0.3011} & 0.3081\\
    \hline
    \end{tabular}
    \end{adjustbox}
    \caption{Results on DIV2K test set (bicubic $\downarrow$4) \cite{Agustsson_2017_CVPR_Workshops}}
    \label{table:bicubic_sr}
\end{table}

\begin{table*}[ht]
    \centering
    \begin{adjustbox}{width=0.85\textwidth}
    \begin{tabular}{|c|c|c|c|c|c|c|}
    \hline
     \multicolumn{7}{|c|}{DIV2K Unknown x4} \\
     \hline
     \multicolumn{4}{|c|}{Mapping}  & \multicolumn{2}{|c|}{} & \multicolumn{1}{|c|}{Mapping x2} \\
     \hline
      & off-the-shelf x2x2 SR & off-the-shelf x4 SR & Specialized x4 SR & EDSR & RCAN & Specialized x2 SR\\
    \hline
    PSNR(dB) & 28.429 & 28.415 & 28.546 & 28.774 & 28.819 & \textbf{28.922}\\
    \hline
    SSIM & 0.813 & 0.811 & 0.817& 0.828 & 0.825& \textbf{0.833}\\
    \hline
    LPIPS-Alex & 0.2977 & 0.2999 & 0.2834 & 0.2797 & 0.2786& \textbf{0.2682}\\
    \hline
    LPIPS-VGG & 0.3342 & 0.3382 & 0.3332 & 0.3109 & 0.3205& \textbf{0.3088}\\
    \hline \hline
    \multicolumn{7}{|c|}{RealSR CANON x4} \\
    \hline
    \multicolumn{4}{|c|}{Mapping} & \multicolumn{2}{|c|}{} & \multicolumn{1}{|c|}{Mapping x2}  \\
    \hline
     & off-the-shelf x2x2 SR & off-the-shelf x4 SR & Specialized x4 SR & EDSR & RCAN & Specialized x2 SR \\
     \hline
    PSNR(dB) & 25.475 & 25.794 & 27.572 & 27.573 & 27.658 & \textbf{27.703} \\
    \hline
    SSIM & 0.754 & 0.762 & 0.794 & 0.792 & 0.793 & \textbf{0.798} \\
    \hline
    LPIPS-Alex & 0.2861 & 0.2920 & 0.2757 & 0.2736 & 0.2797 & \textbf{0.2696}\\
    \hline
    LPIPS-VGG & 0.3905 & 0.3928 & 0.3641 & 0.3637 & \textbf{0.3624} & 0.3649 \\
    \hline \hline
    \multicolumn{7}{|c|}{RealSR NIKON x4} \\
    \hline
    \multicolumn{4}{|c|}{Mapping} & \multicolumn{2}{|c|}{} & \multicolumn{1}{|c|}{Mapping x2} \\
    \hline
     & off-the-shelf x2x2 SR & off-the-shelf x4 SR & Specialized x4 SR & EDSR & RCAN & Specialized x2 SR \\
     \hline
    PSNR(dB) & 23.717 & 24.124 & 26.901 & 26.974 & 27.091 & \textbf{27.107} \\
    \hline
    SSIM & 0.709 & 0.721 & 0.769 & 0.771 & 0.771 & \textbf{0.775}\\
    \hline
    LPIPS-Alex & 0.3296 & 0.3312 & 0.3113 & 0.3093 & 0.3188 & \textbf{0.3043}\\
    \hline
    LPIPS-VGG & 0.4146 & 0.4154 & 0.3839 & 0.3815 & \textbf{0.3801} & 0.3836\\
    \hline
    \end{tabular}
    \end{adjustbox}
    \caption{Test results of the domain mapping diagram \ref{fig:mapping_diagram} for 100 $\downarrow$4 unknown images from DIV2K \cite{Timofte_2018_CVPR_Workshops}, for 50 images from RealSR \cite{caiNTIRE2019} Canon and for 50 images from RealSR Nikon datasets}
    \label{table:mapping_results_sr}
\end{table*}

\begin{table*}[ht]
    \centering
    \begin{adjustbox}{width=0.975\textwidth}
    \begin{tabular}{|c|c|c|c|c|c|c|c|c|c|c|c|c|}
    \hline
     & \multicolumn{6}{|c|}{Canon} & \multicolumn{6}{|c|}{Nikon} \\
    \hline
     & 7Mapped9 & 7x7 & 7Mapped9* & 9x9 & 11Mapped9 & 11x11 & 7Mapped9 & 7x7 & 7Mapped9* & 9x9 & 11Mapped9 & 11x11\\
     \hline
    PSNR(dB) & 39.818 & 20.282 & 40.145 & 40.275 & 37.672 & 24.282 & 37.982 & 19.296 & 38.293 & 38.618 & 35.857 & 23.257 \\
    \hline
    SSIM & 0.972 & 0.578 & 0.974 & 0.975 & 0.959 & 0.676 & 0.9556 & 0.508 & 0.958 & 0.962& 0.932 & 0.615\\
    \hline
    LPIPS-Alex & 0.0469 & 0.3184 & 0.0451 & 0.0411 & 0.0613 & 0.2512 & 0.0732 & 0.3675 & 0.0701 & 0.0606 & 0.0984 & 0.3116\\
    \hline
    LPIPS-VGG & 0.1272 & 0.4061 & 0.1263& 0.1149 & 0.1558 & 0.3646 & 0.1538	& 0.4402 & 0.1545 & 0.1389& 0.1932 & 0.4091\\
     \hline
    \end{tabular}
    \end{adjustbox}
    \caption{Test results using RCAN on RealSR dataset \cite{cai2019toward}- Version 3. 7Mapped9* indicates that mapped images trained as in intermediate stage whereas the other results are only validated after the domain mapping.}
    \label{table:deblurring}
\end{table*}

\section{Evaluation}
\label{sec:experiment}

\subsection{Experimental Setup}
\label{setup}

During training, in each mini-batch, 8 randomly cropped LR patches with size 96×96 are provided as inputs. All models are trained for 150k iterations on a Tesla V100-SXM2 GPU by using ADAM optimizer with $\beta_1= 0.9$ and $\beta_2=0.99$ and $\epsilon= 10^-8$. The learning rate is initialized as $10^-4$ and halved every 50k iterations. For a fair comparison, hyper-parameters are not changed for RCAN \cite{RCAN2018} and EDSR \cite{EDSR2017}. Thus, EDSR contains 8 residual blocks with 64 channels as in RCAN. The training parameters for the domain mapping network are kept the same for SR and image restoration results.

As training set for the SR task, we use 800 LR images from DIV2K unknown dataset \cite{Agustsson_2017_CVPR_Workshops}, in which degradation operators are kept hidden and 400 real-world LR images from RealSR dataset \cite{cai2019toward}. The RealSR dataset contains real LR-HR pairs which are captured by altering the focal length of a camera. To validate the performance of two-stage domain adapted training we use corresponding validation datasets from DIV2K and RealSR. For image quality assessment (IQA) \cite{gu2020image}, we use PSNR and SSIM on RGB channels as distortion metrics, for perception quality evaluation we utilized LPIPS-Alex and LPIPS-VGG \cite{zhang2018perceptual}. 

For the image restoration task, we use 459 images from the RealSR dataset \cite{cai2019toward}. All GT images are degraded by 7x7, 9x9 and 11x11 blur kernels and 40dB noise is added. We select 9x9 blur kernel as the standard degradation model for the intermediate domain. For testing, we used images from Nikon and Canon cameras, which are degraded similarly.  
\vspace{-6pt}

\subsection{Results for Super-Resolution}
\label{result-sr}
First, we wanted to answer the question of what is the best way to perform $\times$4 SR on bicubic downsampled images. Generally, in image super-resolution, 2-consecutive pixel shuffle layers are used to perform $\times$4 SR as proposed in \cite{shi2016realtime}. Even though it is end-to-end trainable, using 2-consecutive pixel shuffle layers is not optimal. Because second pixel shuffle layer can only achieve what is being conveyed from the first pixel shuffle layer. Therefore, as opposed to using direct $\times$4 SR for $\downarrow$4 bicubic LR images, we utilize $\times$2 SR network twice which is trained on DIV2K \cite{Agustsson_2017_CVPR_Workshops} $\downarrow$2 bicubic images $\to$ GT. As shown in table \ref{table:bicubic_sr}, without any additional training, using 2-times $\times$2 SR network performs better than $\times$4 SR network. Therefore, we decide to perform intermediate training for the first results obtained from $\times$2 SR network which leads us to two-stage training that we called specialized $\times$2$\times$2 SR. Results shown in table \ref{table:bicubic_sr} validate us that two-stage domain adapted training for bicubic degraded images performs better than direct $\times$4 and $\times$2$\times$2 SR results. 

The achievement we obtained from domain adapted two-stage training on $\downarrow$4 bicubic LR images drives us to the question of how to perform $\times$4 SR on unknown downsampled images. Starting from the idea proposed in RBSR \cite{Rad2021}, we construct a network to generate mapped $\downarrow$4 bicubic images from unknown $\downarrow$4 LR images. The evaluation results for DIV2K unknown dataset \cite{Timofte_2018_CVPR_Workshops} and RealSR Canon-Nikon dataset \cite{cai2019toward} are shown in table \ref{table:mapping_results_sr}. The best performance for the first branch can be obtained by using specialized two-stage training: first network which is trained on from $\downarrow$4 unknown LR to mapped $\downarrow$4 bicubic LR, then a network to perform SR from those mapped images to $\times$4 SR versions. Yet, direct training from $\downarrow$4 to SR performs better than the first branch of the diagram. The results for direct training with RCAN \cite{RCAN2018} and EDSR \cite{EDSR2017} are shown in the table \ref{table:mapping_results_sr}.

\begin{figure}[htb!]
\begin{minipage}[b]{0.24\linewidth}
  \centering
  \centerline{\includegraphics[width=2cm]{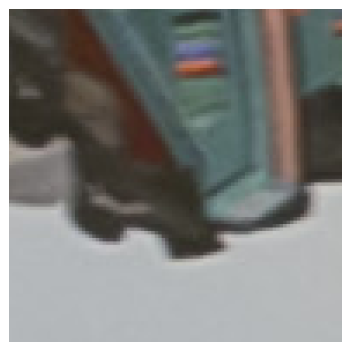}}
  \centerline{(a)}\medskip
\end{minipage}
\begin{minipage}[b]{.24\linewidth}
  \centering
  \centerline{\includegraphics[width=2.0cm]{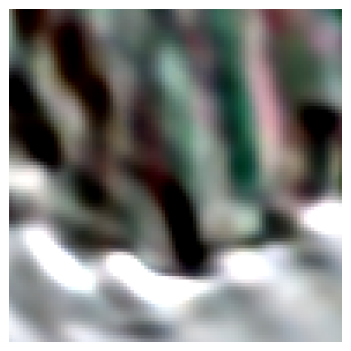}}
  \centerline{(b)}\medskip
\end{minipage}
\hfill
\begin{minipage}[b]{0.24\linewidth}
  \centering
  \centerline{\includegraphics[width=2.0cm]{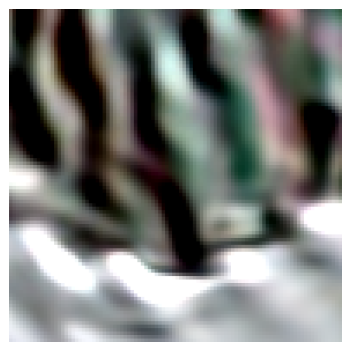}}
  \centerline{(c)}\medskip
\end{minipage}
\begin{minipage}[b]{.24\linewidth}
  \centering
  \centerline{\includegraphics[width=2.0cm]{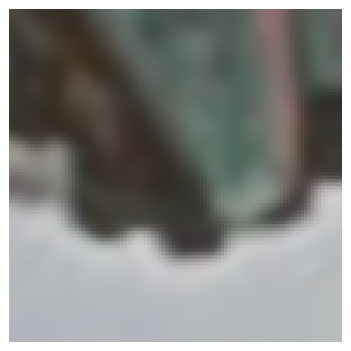}}
  \centerline{(d)}\medskip
\end{minipage} \vspace{-18pt}
\caption{Models are trained on DIV2K unknown dataset \cite{Timofte_2018_CVPR_Workshops} and test on RealSR dataset \cite{cai2019toward}: (a) HR patch Canon 018-PSNR(dB), (b) EDSR-16.813, (c) RCAN-15.575, (d) Mapping $\times$2 and specialized $\times$2 SR-28.145}
\label{fig:canon_test_only}
\end{figure}

However, the performance of image SR models which are trained directly from the unknown degraded LR images with respect to the corresponding GTs are highly dependent on training dataset. Thus, it does not generalize well enough to the other datasets. Because direct training leads to overfitting to the degradation model. Once we change the validation dataset to unknown degraded RealSR \cite{cai2019toward} images, models are overfitted to DIV2K unknown degredation dataset \cite{zhang2018perceptual}, thus they generate artifacts in resulting images which is shown in Fig. \ref{fig:canon_test_only}. However, our proposed two-stage domain adapted training generalizes better compared to the direct training. Therefore, the best recipe for $\downarrow$4 unknown/bicubic images is mapping them to $\downarrow$2 bicubic degraded ones and perform SR from those mapped images. This approach outperforms all other SR approaches when it is evaluated on the same dataset and even when it is trained on a synthetic dataset (DIV2K unknown \cite{Timofte_2018_CVPR_Workshops}) and evaluated on a different dataset with RealSR \cite{cai2019toward} images, it does not produce artifacts and generalizes well compared to other methods.

\vspace{-8mm}

\subsection{Results for Image Restoration}
\label{results-ir}
We choose 9x9 blur as our intermediate domain and we perform domain mapping from different degradation kernels such as 7x7 and 11x11, to the intermediate domain.
Results on Canon and Nikon HR images from the RealSR dataset \cite{cai2019toward} are shown in Table \ref{table:deblurring}. Once we map 7x7 blur or 11x11 to 9x9 blur, a model trained on 9x9 blurred provides improvements up to 19dB compared to a network trained on 9x9 blurred images applied directly to 7x7 or 11x11 blurred images. Once 7-Mapped-9 images are obtained, we perform specialized deblurring by using those mapped images as an intermediate training and 7-Mapped-9* column validates how successful two-stage domain adapted training is.   

\vspace{-2ex}

\section{Conclusion} \vspace{-6pt}
\label{sec:conclude}
In this work, we have shown how to perform $\times$4 SR effectively when degradation model is unknown by proposing a two-stage domain adapted training approach. This idea is also applicable to other real-world ill-posed image restoration problems. Therefore, it is better to map given images from an unknown space to a tractable one and perform restoration/SR to overcome overfitting to the degradation model. We show that our approach, two-stage domain adapted training, outperforms direct training in real-world restoration/SR both qualitatively and quantitatively.


\clearpage

\bibliographystyle{IEEEbib}
\bibliography{strings,refs}

\end{document}